\documentclass[fleqn,10pt]{wlscirep}
\title{Fast atom transport and launching in a nonrigid trap}

\newcommand{\beq}{\begin{equation}}
\newcommand{\eeq}{\end{equation}}
\newcommand{\beqa}{\begin{eqnarray}}
\newcommand{\eeqa}{\end{eqnarray}}
\usepackage{calrsfs}
\DeclareMathAlphabet{\pazocal}{OMS}{zplm}{m}{n}

\author{A. Tobalina, M. Palmero, S. Mart\'inez-Garaot, and J. G. Muga}

\affil{Departamento de Qu\'{\i}mica F\'{\i}sica, UPV/EHU, Apartado 644, 48080 Bilbao, Spain}
\begin{abstract}
We study the shuttling  of an atom in a trap with controllable position and  frequency. 
Using invariant-based inverse engineering, protocols in which the trap is simultaneously displaced and expanded are proposed
to speed up transport between stationary trap locations as well as  launching processes
with narrow final-velocity distributions.  
Depending on the physical constraints imposed, either simultaneous or sequential approaches may be faster.   
We consider first a perfectly harmonic trap, and then extend the treatment to generic traps. 
Finally, we apply this general framework to a double-well potential to separate different motional states
with different launching velocities.
\end{abstract}

\begin{document}

\flushbottom
\maketitle
\thispagestyle{empty}
\section*{Introduction}
\begin{figure}[b]
\begin{center}
\includegraphics[width=0.7 \linewidth]{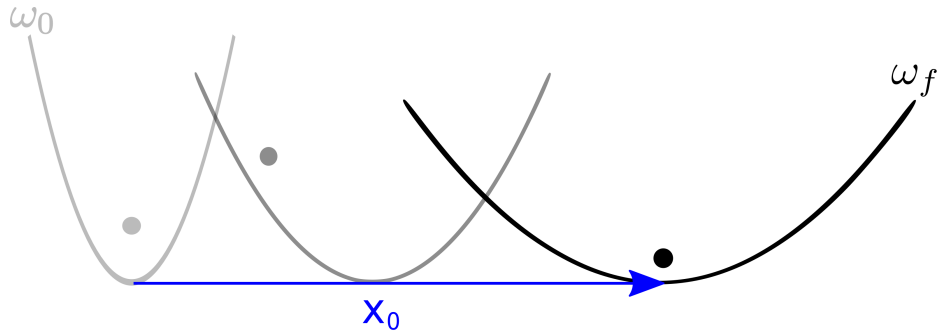}
\caption{\label{scheme1}(Color online)
Scheme of the transport protocol with a change in the frequency of the trap.}
\end{center}
\end{figure}

An important  goal of modern atomic physics is to control  atomic motion for fundamental studies or 
to develop quantum-based technologies. 
%Cooling up to ultracold temperatures is an  example
%of this work.
%, where achieving  low average kinetic energy  is the objective. 
Technological advances 
allow for driving individual atoms (ions \cite{Bowler2012,Walther2012} or neutral atoms \cite{Steffen2012}) 
along microscopic or mesoscopic predetermined 
space-time paths. 
This control will enable us to use the rich structure and interactions of ions and neutral atoms 
in circuits and devices where quantum phenomena play a significant role. Many operations require 
moving the atoms fast to keep quantum coherence,  leaving them unexcited at their destination. 
Slow adiabatic shuttling may avoid excitation  in principle, but the long times required make the processes  prone to decoherence.   
Shortcuts to adiabaticity (STA) \cite{Chen2010a,Torrontegui2013} are protocols for the control parameters that produce final states of an 
adiabatic process in much shorter times, typically via diabatic transitions at intermediate times. 
In this paper, we find STA to drive   a single atom  by a moving and nonrigid potential
with time-dependent frequency as schematically shown in Fig. \ref{scheme1}. 
We shall focus first on harmonic traps, and then a theory for more general potentials is also put forward.  
% % % % % % % % % % % % % % % % % % % % % % % % % % % % % % % % % % % % % % % % %
% % % % % % % % % % % % % % % % % % % % % % % % % % % % % % % % % % % % % % % %
% % % % % % % % % % % % % % % % % % % % % % % % % % % % % % % % % % % % % % % % % % %
% % % % % % % % % % % % % % % % % % % % % % % % % % % % % % % % % % % % % % % % % % %
%Shuttling of a double-well potential is also considered for being a particularly interesting potential.
Two types of basic processes addressed are: ({\it i}) transport where the wave packet center and trap start and end at rest, and also ({\it ii}) launching or stopping processes, 
where the wave-packet center and trap start (resp. end) at rest, and ends (resp. start) with a  nonzero velocity. 
Invariant-based inverse engineering has been applied to designing STA for rigid transport (with a constant potential in the moving frame)  \cite{Torrontegui2011,Chen2011,Furst2014}, and trap expansions or compressions   \cite{Chen2010a,Chen2010b,Stefanatos2010,Torrontegui2012a,Cui2016}. 
While shuttling and expansion or compression could be performed sequentially, doing both operations simultaneously, 
as proposed here, may save time and offers broader control possibilities.  
``Dual-task'' operations must thus be compared to sequential operations. 
In principle, STA for rigid transport and expansions can be done in arbitrarily short times, but only if infinite resources and energies are available, 
which is never the case in practice. 
Often, the control parameters cannot go beyond certain values. For example, a very fast trap expansion without final excitation 
needs transient imaginary frequencies of the external trap (a concave-down potential), 
which are not easy to implement in all trap types. 
In optical traps, for example, the passage through the atomic resonance of the laser frequency to go from a trap to an antitrap may produce undesired excitation. A different, common  constraint is the limitation on the 
spatial domain allowed for the trap center.  
We shall show that, depending on the constraints imposed,  either sequential or dual-task protocols 
may be faster.

There are different fields or applications where simultaneous transport and expansion or compression between initial and final states at rest is  of relevance.  
In quantum heat engines and refrigerators \cite{Rezek2006,Salamon2009,Hoffmann2011,Abah2012, Deng2013,Jarzynski2013,Stefanatos2014,DelCampo2014,Beau2016,Rossnagel2016}
for example, the (thermodynamically) adiabatic expansion or compression strokes 
of the cycle could be realized simultaneously transporting the quantum working medium between baths at different locations.  
Also, when expanding or separating ion chains, which are basic processes to develop  a scalable quantum-information architecture \cite{Wineland1998}, 
the effective dynamics of the normal modes involves simultaneous transport and frequency change  
\cite{Palmero2015,Palmero2015a}.  One more scenario where transport and frequency change occur simultaneously is the bias inversion of an asymmetric  double-well potential \cite{Martinez-Garaot2015}. 

Launching and stopping protocols are as well useful for many applications. 
An example of a stopping device is the ``inverse coil gun'' implemented by Mark Raizen and coworkers \cite{Narevicius2008}. 
It uses pulsed magnetic fields to slow down a supersonic beam  
(e.g. from $500$ to $50$ m/s \cite{Narevicius2008}) so as to leave the atoms ready for spectroscopic studies, controlled collisions, or further cooling techniques. 
One advantage of stopping techniques by magnetic (for paramagnetic species) or electric fields (for ions), 
is their broad range of applicability,  beyond the very restricted class of atoms with a  cycling transition that can be treated 
by standard laser cooling approaches. 
The opposite process, launching, is also of much current interest:  launching ions with a specific speed is used in particular for their implantation or deposition
\cite{Jacob2016}.
Accurately controlled launching can contribute to different quantum technologies such as ion microscopy, those using a controlled ``soft landing'' of slow ions on a surface, 
and those controlling the location of defects (NV centers) that have been proposed for sensors and also as the basis of a possible
architectures for quantum information processing. Deterministic sources of single cold ions have been proposed and demonstrated 
\cite{Meijer2006,Jacob2016} that limit the position-momentum uncertainty only due to the Heisenberg principle. 
Our goal here  is to control of the velocity, and its dispersion.  
%more accurate than with a standard voltage ramp. 
This is facilitated by the possibility to change the trap frequency along the shuttling.    
Differential launching of different motional states is also possible as we shall demonstrate with a double well.  

While the mathematical framework of this work is equally applicable to neutral atoms or trapped ions, 
the numerical examples make use of parameters adapted to trapped ions \cite{Bowler2012,Walther2012}. 
%
%
%
%\section*{Results}
%
%
%
%
%
%
\subsection*{Invariant-based inverse engineering}
Lewis and Riesenfeld \cite{Lewis1969} noted that the solutions of the Schr\" odinger equation 
for a time-dependent Hamiltonian can be written  as superpositions %with constant coefficients% 
of eigenstates of its dynamical invariants.  
Dhara and Lawande \cite{Dhara1984} and Lewis and Leach \cite{Lewis1982} worked out the details for a particle of mass $m$ that evolves according to Hamiltonians of the form%moving in potentials of the form
\beq
%\label{LLpotential}
\label{LLHamiltonian}
%V = -F(t)x + \frac{m}{2}\omega ^2(t)x^2 + \frac{1}{\rho ^2(t)}U\left[\frac{x-\alpha (t)}{\rho (t)}\right],
H=\frac{p^2}{2m} -F(t)x + \frac{m}{2}\omega ^2(t)x^2 + \frac{1}{\rho ^2(t)}U\left[\frac{x-\alpha (t)}{\rho (t)}\right], 
\eeq
where $F(t)$ is a homogeneous force, $\omega (t)/(2\pi)$ the frequency of a harmonic term, $U$ an arbitrary function, and $\alpha(t)$ and  $\rho (t)$ are auxiliary functions.  $x$ and $p$ represent conjugate position and momentum 
operators of the particle. 
  
%The corresponding Hamiltonian $H=p^2/(2m)+V$ has the quadratic-in-momentum invariant 
The Hamiltonian in Eq. (\ref{LLHamiltonian}) has the quadratic-in-momentum invariant
\beq
\label{invariant}
I = \frac{1}{2m} \left[ \rho (p-m\dot{\alpha}) - m\dot{\rho}(x-\alpha) \right]^2+ \frac{1}{2}m\omega_0^2 \left( \frac{x-\alpha}{\rho} \right)^2 + U \left( \frac{x-\alpha}{\rho} \right),
\eeq
where the dot means time derivative.  
$I$ satisfies indeed the invariance equation
\beq
\label{invequation}
\frac{dI}{dt} \equiv \frac{\partial I(t)}{\partial t} + \frac{1}{i\hbar} [I(t),H(t)]=0,
\eeq
provided the scaling factor $\rho$ and $\alpha$ satisfy the Ermakov and Newton equations,
\beqa
\label{ermakov}
\ddot{\rho} + \omega ^2(t) \rho &=& \frac{\omega_0^2}{\rho ^3},
\\
\label{newton}
\ddot{\alpha} + \omega ^2(t) \alpha &=& \frac{F(t)}{m},
\eeqa
where $\omega_0$ is a constant. For simplicity we choose $\omega_0=\omega (0)$.

Any wavefunction $\psi(t)$ driven by the Hamiltonian (\ref{LLHamiltonian})
may be written in terms of eigenvectors  $\psi_n$ of the invariant (\ref{invariant}),
\beqa
\label{eigenvalueequation}
\psi (x,t) = \sum_n c_ne^{i\theta_n} \psi_n (x,t),\;\;\;\;\;
%\nonumber\\
I(t)\psi_n (x,t) = \lambda_n\psi_n(x,t),
\eeqa
where $c_n$ are constant coefficients, the $\lambda_n$ are the eigenvalues, and $\theta_n$ are 
Lewis-Riesenfeld phases that can be calculated from $H$ and $\psi_n$ \cite{Lewis1969}, 
$
\theta_n(t)=\frac{1}{\hbar}\int_0^{t_f} dt' \langle \psi_n(t')|i\hbar \frac{\partial}{\partial t}-H(t')|\psi_n(t')\rangle.
$
The $\psi_n$  have the form \cite{Dhara1984}
\beq
\label{eigenvecinv}
\psi_n(x,t) = e^{\frac{im}{\hbar} \left[ \dot{\rho}x^2/2\rho + (\dot{\alpha}\rho - \alpha \dot{\rho})x/\rho \right]} \frac{1}{\rho^{1/2}} \phi_n \left( \frac{x-\alpha}{\rho} \right),
\eeq
where the $\phi_n(\sigma)$ (normalized in $\sigma := \frac{x-\alpha}{\rho}$ space) are the solutions of the auxiliary, stationary Schr\"odinger equation 
%for the static potential $1/2m\omega_0^2\sigma^2+U(\sigma)$.  
% (normalized in $\sigma := \frac{x-\alpha}{\rho}$ *********this cannot be general**********: 
%
\beq
\label{auxschrodinger}
\left[-\frac{\hbar^2}{2m} \frac{\partial^2}{\partial \sigma ^2} + \frac 1 2 m \omega_0^2 \sigma^2 + U(\sigma)\right]\phi_n(\sigma)=\lambda_n\phi_n(\sigma).
\eeq
The physical meaning of $\alpha$ is made evident in Eq. (\ref{eigenvecinv}) as a centroid for the dynamical wavefunctions that satisfies the Newton equation (\ref{newton}).  $\alpha$  is also the center of the potential term $\rho^{-2}U[(x-\alpha)/\rho]$ when $U$ does not vanish. 

To inverse engineer the interaction between the initial time, $t=0$, and a final time $t_f$, we first set the initial and final Hamiltonians. 
For transport between stationary traps, commutativity is imposed between the Hamiltonian and the invariant at boundary times so that they share eigenstates. Thus the dynamics maps eigenstates of $H(0)$ onto eigenstates of $H(t_f)$ via the corresponding invariant eigenstates, even though at intermediate times diabatic transitions may occur. 
%prove that in a reference frame that moves and expands with the trap, the Hamiltonian and the invariant share eigenstates at all times.
The commutation of $H$ and $I$ at boundary times implies boundary conditions for 
$\alpha$, $\rho$, and their derivatives. We design these functions to satisfy the necessary boundary conditions, 
and then, from the auxiliary Eqs. (\ref{ermakov}) and (\ref{newton}) the control parameters $\omega(t)$ and $F(t)$ are found. For launching/stopping processes the invariant and Hamiltonian do not commute at final time in the laboratory frame, but the states may be chosen as   eigenstates of the Hamiltonian in the comoving and coexpanding frame.    
\section*{Results}
\subsection*{Dual-task transport in a nonrigid harmonic trap}
Let us assume first that the external trap is purely harmonic, i.e., we take $U=0$ and $F=m\omega^2 (t) x_0(t)$, where $x_0(t)$ is the position of the trap center. 
Then, the Hamiltonian in Eq. (\ref{LLHamiltonian}) becomes, adding  a purely time-dependent term that does not affect the physics to complete the square,   
\beq
\label{Hamiltonian}
H = \frac{p^2}{2m} + \frac{1}{2}m\omega^2 (t) [x-x_0(t)]^2.
\eeq
The average energy for this system in the $n$th state (\ref{eigenvecinv}) is given by 
\beqa
\label{stateenergy}
E = \langle H \rangle = \frac{(2n+1)\hbar}{4\omega_0} \left[ \dot{\rho}^2 +\omega^2 (t) \rho^2 + \frac{\omega_0^2}{\rho^2} \right]
+ \frac{1}{2}m\dot{\alpha} + \frac{1}{2}m\omega^2 (t) [\alpha-x_0 (t)]^2.
\eeqa
%
%Details on how to obtain this expression are shown in \ref{AppendixA}.
For rigid transport \cite{Torrontegui2011},  $\omega$ is constant and Eq. (\ref{ermakov}) is trivially satisfied for $\rho (t)=1$. 
Here, the goal is to transport a particle a distance $d$, and additionally change the angular frequency of the trap from the initial value  $\omega _0$ to the final value $\omega_f\equiv\omega(t_f)=\omega_0/\gamma^2$, without final excitation. 
The control parameters are the frequency $\omega (t)$ and the position of the center of the trap $x_0 (t)$. 
Figure \ref{scheme1} shows schematically this process. 
The auxiliary functions $\alpha(t)$ and $\rho(t)$ have to satisfy the boundary conditions
\beqa
\label{bc1}
\alpha(0)&=&0,\quad \alpha(t_f)=d,
\nonumber\\
\rho(0)&=&1,\quad \rho(t_f)=\gamma.
\eeqa
%
%Note that, for a harmonic trap, the meaning of the auxiliary parameter $\alpha(t)$ is not the center of $U$, we have $U=0$, but rather 
%a classical trajectory that the states $\psi_n$ follow.    
We also find, by imposing commutativity between Hamiltonian and invariant at boundary times, the boundary conditions
\beqa
\label{bc2}
\dot{\alpha}(0)&=&\dot{\alpha}(t_f)=0,
\nonumber\\
\dot{\rho}(0)&=&\dot{\rho}(t_f)=0. 
\eeqa
Additionally, to satisfy the invariant condition in Eq. (\ref{invequation}) we need to impose
\beqa
\label{bc3}
\ddot{\alpha}(0) &=& \ddot{\alpha}(t_f) =0,
\nonumber\\
\ddot{\rho}(0) &=& \ddot{\rho}(t_f) =0.
\eeqa
Now, we may propose ansatzes that satisfy all boundary conditions in Eqs. (\ref{bc1}), (\ref{bc2}) and (\ref{bc3}). A simple choice is $\rho (t) = \sum_{i=0}^5 \rho_i s^i$ and $\alpha (t) = \sum_{i=0}^5 \alpha_i s^i$, where $s=t/t_f$. Fixing the coefficients $\rho_i$ and $\alpha_i$ to satisfy the boundary conditions, the auxiliary functions become 
\beqa
\label{ansatz}
\rho (t) = 1+10(\gamma -1) s^3 - 15(\gamma -1)s^4 +6(\gamma -1)s^5,\;\;\;\;\;\;
%\nonumber\\
\alpha (t) = 10ds^3 - 15ds^4 + 6ds^5.
\eeqa
Substituting $\rho$ in Eq. (\ref{ermakov}), the time dependent frequency in (\ref{Hamiltonian}) takes the form 
\beq
\label{omega}
\omega (t) = \sqrt{\frac{\omega_0^2}{\rho^4}-\frac{\ddot{\rho}}{\rho}}, 
\eeq
whereas, from Eq. (\ref{newton}),  the transport function (position of the trap center) is 
\beq
\label{position}
x_0 (t) = \frac{\ddot{\alpha}}{\omega ^2}+\alpha,
\eeq
that can be now calculated with Eqs. (\ref{ansatz}) and (\ref{omega}). The form of the polynomial for $\rho$ in Eq. (\ref{ansatz}) is not affected by the transport, 
so the function for the frequency in Eq. (\ref{omega}) is the same as the one used for pure expansions  \cite{Chen2010a}. 
Similarly, the form of $\alpha(t)$ is not affected by the expansion, but the trap position $x_0(t)$ 
is different from the one in rigid transport \cite{Torrontegui2011} due to the time dependence of the frequency. 
The dual task protocol is thus not just a simultaneous superposition of recipes for pure expansions and rigid transport but a genuinely different process. 
%-----------------------------------------
%-----------------------------------------
\begin{table}[t]
\centering
\begin{tabular}{|l|c|c|c|}
\hline
 &$\omega>0$ & trap in $[0,d]$ &Both conditions\\ 
 \hline
Sequential\hspace{0.5cm} &0.443 $\mu$s&0.2 $\mu$s&0.643 $\mu$s\\
\hline
Dual\hspace{0.5cm} &0.443 $\mu$s&0.91 $\mu$s  &0.91 $\mu$s\\
\hline
\end{tabular}
\caption{\label{tab_trans} Minimal times for the transport+expansion process when the trap frequency or/and center are limited, see text.
Parameters: $d=370$ $\mu$m, $\gamma=\sqrt{10}$, and  $\omega_0/(2 \pi)=2$ MHz.}
\end{table}
%-----------------------------------------
%-----------------------------------------

%
%%-----------------------------------------
%%-----------------------------------------
%\begin{table}
%\caption{\label{tab_trans} Optimal final times for the transport+expansion process.}
%\bigskip
%\footnotesize
%\begin{center}
%\begin{tabular}{@{}llll}
%\br
% &Expansion time (real $\omega$) &Transport time (in box) &Total time\\
%\mr
%Sequential\hspace{0.5cm} &0.443 $\mu$s&0.2 $\mu$s&0.643 $\mu$s\\
%Dual\hspace{0.5cm} &0.443 $\mu$s&0.91 $\mu$s&0.91 $\mu$s\\
%\br
%\end{tabular}
%\end{center}
%\end{table}
%\normalsize
%%-----------------------------------------
%%-----------------------------------------

%the frequency and is means that intrinsically, it is different to perform the transport plus the expansion/compression sequentially, 
%combining the simple transport and simple expansion/compression, 
%or to do it via the dual protocol designed before. 
We performed a number of tests to compare the times required by the sequential or dual protocols.
In principle, both the sequential and the dual drivings can be done arbitrarily fast, if no limitations are imposed. 
However, subjected to technical limitations the minimal times may be different. 
One of the bounds will be to keep the frequency always real, $\omega^2(t)>0$, since  a repulsive parabola may be
difficult to implement in some trapping methods.  
Other natural constraint is to limit the trap position bounded within the ``box'' $[0,d]$.
%Exceeding this ``box'' would require a larger space, namely, a larger trap so that the atom can be moved without actually leaving the trapping zone.  

We carry out the comparisons for a $^9$Be$^+$ ion, shuttled over a distance $d=370$ $\mu$m in a trap with
initial frequency $\omega_0/(2 \pi)=2$ MHz expanded by a factor of 10, $\gamma^2=10$.
For these parameters and polynomial ansatzes, the simple expansion has a minimal final time $t_{f_{exp}}^{(min)}=0.443$ $\mu$s, below which imaginary frequencies appear.  
Note that this will also be the limit time before getting imaginary frequencies in the dual process, as Eq. (\ref{omega}) gives exactly the same evolution for $\omega$ in a simple expansion or a dual process.
For rigid transport, carried out before the expansion at the highest trap frequency, 
the limit time is $t_{f_{tra}}^{(min)}=0.2$ $\mu$s before exceeding the box. 
Thus, the total minimal time for the sequential protocol is $t_{f_{seq}}=0.643$ $\mu$s.
For the dual protocol, the minimal final time before exceeding the box is $t_{f_{dual}}=0.91$ $\mu$s. 
%larger than the minimal total sequential time.
Under the stated restrictions (real frequencies and the trap bounded by the predetermined box $[0,d]$), 
the dual protocol is slower than the sequential one, 
if performing the transport first and then the expansion.
All final times are summarized in Table \ref{tab_trans}.

\begin{figure}[t]
\begin{center}
\includegraphics[width=0.7 \linewidth]{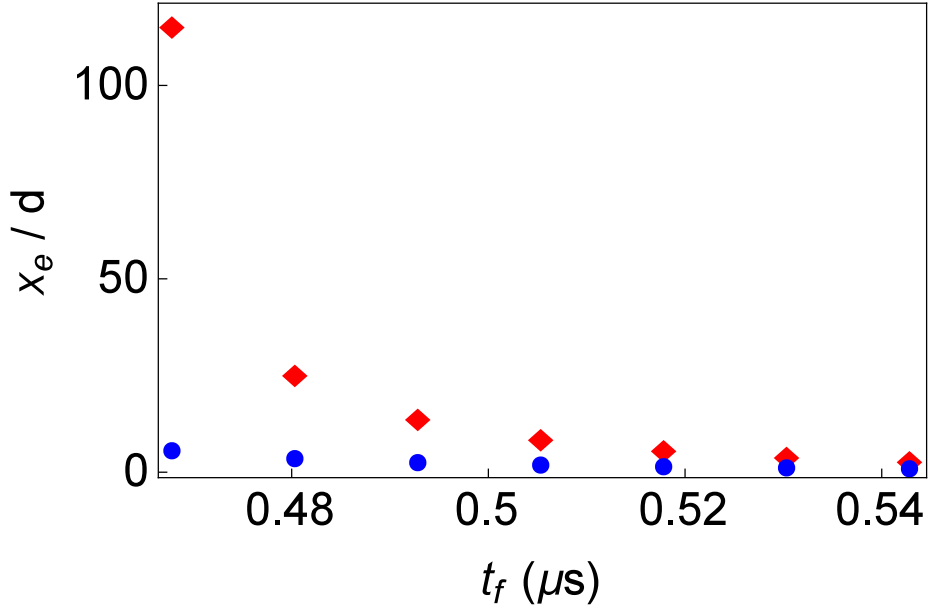}
\caption{\label{exceeded}(Color online)
Ratio of the exceeded distance $x_e$ and the transport distance $d$ for the dual (blue circles) and sequential (red diamonds) non-rigid harmonic tranport protocols,
for final times that do not require imaginary frequencies. Parameters used are $d=370$ $\mu$m, $\gamma=\sqrt{10}$, and  $\omega_0/(2 \pi)=2$ MHz.}
\end{center}
\end{figure}

If the only restriction is to keep real frequencies, dropping the limitation on the domain of the trap position,    
the minimal final time is in principle $t_f^{(min)}=0.443$ $\mu$s for both the sequential and dual protocols, but 
in the sequential protocol this is a really challenging limit since the transport should be done in  zero time.   
In both protocols the transport function exceeds the box $[0,d]$.   
In Fig. \ref{exceeded} we compare the ratio between the exceeded distance beyond $[0,d]$   
and $d$ for the sequential and the dual drivings, with respect to the total process time. The exceeded distance is defined in terms of the maximum ($x_{0_{max}}$) and the minimum ($x_{0_{min}}$) values of the trajectory as $x_e=x_{0_{max}}-x_{0_{min}}-d$.
The figure shows that the dual protocol is much more robust. 
As the minimal possible time is approached, the ratio in the sequential protocol increases dramatically.
%, implying a prohibitive domain for the trap motion. 
In contrast, the ratio in the dual protocol is very stable, making potentially easier to perform the dual protocol for short times.  

% % % % % % % % % % % % % % % % % % % % % % % % % % % % % % % % % % % % % % % % %
% % % % % % % % % % % % % % % % % % % % % % % % % % % % % % % % % % % % % % % %
% % % % % % % % % % % % % % % % % % % % % % % % % % % % % % % % % % % % % % % % % % %
% % % % % % % % % % % % % % % % % % % % % % % % % % % % % % % % % % % % % % % % % % %

%
%
%
\subsection*{Dual-task launching in a harmonic trap}
We study now launching processes where the frequency of the trap is time dependent
(stopping processes may be designed by inverting the launching protocols).  
If the ion is to be launched adiabatically with a very precise velocity, the trap should have a small final frequency to minimize the uncertainty. STA protocols will achieve the same goal in a shorter time.

The order of the sequence plays a relevant role to compare sequential or dual launching protocols. 
In the previous subsection, when the final state is at rest,  the sequential protocol may be  faster than the dual one when transport is done first, then the expansion.  
For the launching process, the only meaningful sequential process implies to expand first, and then to transport,
%the expansion (at the final position) cannot be done after the transport, 
%as the it must have a nonzero final velocity%in order to launch the state
but a small trap frequency does not enable us to implement a fast launching.
%The reverse order is possible but it implies transport with lower frequency. 
%As a consequence a dual launching protocol can be done in shorter times than the sequential expansion (first) + transport (later) process.  
It is therefore useful to combine the time dependences of frequency and displacement of
the trap in a dual protocol. 

The boundary conditions to be imposed for this launching protocol are the same as in Eqs. (\ref{bc1}), (\ref{bc2}) and (\ref{bc3}), except 
that the first derivative of $\alpha$ at final time, 
is now the final launching velocity $v_f$,  
\beq
\label{bc4}
\dot{\alpha}(t_f)=v_f. 
\eeq
Additionally, boundary conditions are imposed on the third derivative of $\alpha$,
\beq
\label{bc5}
\alpha^{(3)}(0)=\alpha^{(3)}(t_f)=0,
\eeq
where $^{(n)}$ means $n$th derivative, so that, according to Eq. (\ref{position}), 
the velocity of the trap $\dot{x}_0$ and the velocity of the wave packet $\dot{\alpha}$ are the same at the boundary times. In order to satisfy the additional boundary conditions, we consider a higher-order polynomial ansatz for $\alpha$, $\alpha=\sum_{i=0}^7\alpha_is^i$, which upon fixing parameters to satisfy all boundary conditions gives
\beqa
\label{ansatzalphalaunching}
\alpha(t)=5(7d-3t_f v_f)s^4-3(28d-13t_fv_f)s^5
+2(35d-17t_fv_f)s^6-10(2d-t_fv_f)s^7.
\eeqa
Boundary conditions for $\rho$ are the same as in the previous subsection, so the same ansatz used in Eq. (\ref{ansatz}) is valid here.  
Thus, the evolution of the frequency is given in Eq. (\ref{omega}), while the evolution of the trap position is found substituting
Eqs. (\ref{omega}) and (\ref{ansatzalphalaunching}) into Eq. (\ref{position}).

We evaluated the sequential and dual launching protocols limiting the frequencies to real values and the domain of the trap center to $[0,d]$. 
For the same parameters used in the previous subsection, and for a final velocity $v_f=10$ m/s, 
%we evaluate launching protocols so that they
%neither have imaginary frequencies, nor exceed the box at any time during the process. 
the minimal expansion time is the one given in the previous subsection, $t_{f_{exp}}^{(min)}=0.443$ $\mu$s, as the expansion  does not change for the new boundary conditions.
The rigid transport, however, performed with the final trap frequency, can be done in a minimal time 
$t_{f_{tra}}^{(min)}=2.295$ $\mu$s without  exceeding the box. 
Thus, the minimal sequential time is $t_{f{tot}}^{(min)}=2.734$ $\mu$s.
For the dual protocol, the minimal  time not  exceeding the box is $t_{f_{dual}}=1.216$ $\mu$s.
The times are summarized in Table \ref{tab_launch}.
Here the dual protocol clearly outperforms the sequential one. 
%

%-----------------------------------------
%-----------------------------------------
\begin{table}[t]
\centering
\begin{tabular}{|l|c|c|c|}
\hline
 & $\omega>0$ &trap in $[0,d]$&Both conditions\\ 
 \hline
Sequential\hspace{0.5cm} &0.443 $\mu$s&2.295 $\mu$s&2.734 $\mu$s\\
\hline
Dual\hspace{0.5cm} &0.443 $\mu$s&1.216 $\mu$s &1.216 $\mu$s\\
\hline
\end{tabular}
\caption{\label{tab_launch} Minimal final times for the launching+expansion process with limited frequency or/and trap center, see text.
Parameters: $d=370$ $\mu$m, $\gamma=\sqrt{10}$, $v_f=10$ m/s, and  $\omega_0/(2 \pi)=2$ MHz.}
\end{table}
%-----------------------------------------
%-----------------------------------------

%%-----------------------------------------
%%-----------------------------------------
%\begin{table}
%\caption{\label{tab_launch} Minimal final times for the launching+expansion process.}
%\bigskip
%\footnotesize
%\begin{center}
%\begin{tabular}{@{}llll}
%\br
%  &Expansion time (real $\omega$) &Transport time (in box) &Total time\\
%\mr
%Sequential\hspace{0.5cm} &0.443 $\mu$s&2.295 $\mu$s&2.734 $\mu$s\\
%Dual\hspace{0.5cm} &0.443 $\mu$s&1.216 $\mu$s&1.216 $\mu$s\\
%\br
%\end{tabular}
%\end{center}
%\end{table}
%\normalsize
%%-----------------------------------------
%%-----------------------------------------

A control possibility we have for the dual process, which does not exist for the sequential one, is to design the launching 
with a given constant expanding velocity, i.e., we impose $\dot{\alpha}(t_f)=v_f$ as before  
and also 
%For that end we change the boundary condition for $\dot{\rho}(t_f)$ at final time as we already did for $\dot{\alpha}$ in Eq. (\ref{bc4}), to  
%The new boundary condition will be
%
\beq
\label{bc6}
\dot{\rho}(t_f)=\epsilon.
\eeq
Additionally, boundary conditions may be imposed on the third derivative,  
\beq
\label{bc7}
\rho^{(3)}(0)=\rho^{(3)}(t_f)=0,
\eeq
so that, from Eq. (\ref{ermakov}),
$\dot{\omega}(0)=0$ and $\dot{\omega}(t_f)=-2\epsilon\omega_0/\gamma^3$, which guarantees that the expansion velocity of the dynamical state matches 
that of the instantaneous eigenstates of the trap,  consistently with the time derivative of $\rho(t_f)=\sqrt{\frac{\omega_0}{\omega_f}}$. 
%to guarantee 
%that the final expanding velocity of the trap matches that of the wave packet,

For the polynomial ansatz  
$\rho=\sum_{i=0}^7\rho_is^i$ 
the coefficients are fixed to satisfy the boundary conditions,  
\beqa
\label{rholaunch}
\rho(t)=1+5(-7+7\gamma-3\epsilon t_f)s^4-3(-28+28\gamma-13\epsilon t_f)s^5
%\nonumber\\
+2(-35+35\gamma-17\epsilon t_f)s^6-10(-2+2\gamma-\epsilon t_f)s^7.
\eeqa
%
%This time dependence of the auxiliary parameter $\rho(t)$ leads to 
%
With the evolutions considered in this section, either for the expanding or the nonexpanding launching, 
a state which is initially an eigenstate of $H(0)$ will not become an eigenstate of the Hamiltonian $H(t_f)$. 
Instead, the state of the system at the end of the process is, see Eq. (\ref{eigenvecinv}),  
$
\psi_n(x,t_f) = e^{\frac{im}{\hbar} \left[ \epsilon x^2/2\gamma + (v_f\gamma - d \epsilon)x/\gamma \right]} \frac{1}{\gamma^{1/2}} \phi_n \left( \frac{x-d}{\gamma} \right),
$
which can be shown to correspond to the Hamiltonian 
eigenstate in the moving and expanding reference system of the trap (see Methods).

The expectation value of the velocity for $\psi_n(x,t_f)$ is $v_f$ and its dispersion is  
\beqa
\label{disp0}
\Delta v = \sqrt{\frac{\hbar(2n+1)}{2m\omega_0}\left(\gamma^2\epsilon^2+\frac{\omega_0^2}{\gamma^4}\right)},
\eeqa
minimal with respect to $\epsilon$ for $\epsilon=0$.  
It  can be lowered further by  decreasing the final trap frequency (increasing $\gamma$).  
This result may be compared with the process where the initial trap is turned off and
a constant electric field is applied. Then the  
dispersion does not change,     
$
\Delta v = \sqrt{[\hbar(2n+1)\omega_0]/(2m)}. 
$
Much smaller  spreads can be achieved by the dual protocol, but $\gamma$ cannot be made arbitrarily small 
in a fixed process time. In particular, the requirement of keeping  the frequency real implies the bound \cite{Salamon2009,Chen2010b} 
$t_f>\sqrt{\gamma^2-1}/\omega_0$. A constant electric field  has its own, different limitations, 
in particular, with constant acceleration  the time is fixed as  $t_f=2d/v_f$
to reach a given final velocity $v_f$ in a distance $d$.    
\subsection*{Dual-task shortcuts in an arbitrary trap}
Now, we extend the analysis to move and expand or compress an arbitrary confining potential from $U(x)$ to 
$\frac{1}{\rho(t_f)^2}U \left[ \frac{x-\alpha(t_f)}{\rho(t_f)} \right]$.    
To stay within the family of processes described by Eq. (\ref{LLHamiltonian}), so that invariants are known, we must impose 
that the harmonic and linear terms depending on $\omega^2$ and $F$ vanish 
at the boundary times. We thus set $\omega_0=0$ hereafter. If initial and final
potentials  are at rest, by imposing commutativity between the Hamiltonian (\ref{LLHamiltonian}) and the invariant (\ref{invariant}) and continuity at the boundary times, we get the same boundary conditions as in Eqs. (\ref{bc2}) and (\ref{bc3}). We must also impose the boundary conditions in Eq. (\ref{bc1}) for the system to be displaced and expanded or compressed, noting that now the constant $\gamma$ is not related to $\omega_0$. 
With these boundary conditions, using  the auxiliary Eqs. (\ref{ermakov}) and (\ref{newton}),  $F(0)=F(t_f)=\omega(0)=\omega(t_f)=0$. That is, the only non vanishing term of the potential  at the boundary times $t_b=0, t_f$  is 
$
%\label{potentialU}
V(t_b)=\frac{1}{\rho(t_b)^2}U\left[ \frac{x-\alpha(t_b)}{\rho(t_b)} \right].
$
We design the functions $\alpha(t)$ and $\rho(t)$ polynomially as before, so that they satisfy all boundary conditions, and introduce them in the auxiliary equations to inversely obtain the control parameters. The auxiliary functions can be the same as in Eq. (\ref{ansatz}). Substituting $\rho$ in Eq. (\ref{ermakov}), 
\beq
\label{auxiliaryharmonic}
\omega^2(t) = -\frac{\ddot{\rho}}{\rho},
\eeq
and substituting this result and $\alpha$ in Eq. (\ref{newton}) we get
\beq
\label{auxiliarylinear}
F(t)=m\ddot{\alpha}+m\omega^2\alpha.
\eeq
In other words, the protocol requires auxiliary time-dependent linear and quadratic potential terms apart from the scaled 
potential $\frac{1}{\rho^2(t)}U[\frac{x-\alpha(t)}{\rho}]$. 
This protocol is of course technically more demanding than the one designed for the simple harmonic trap, because of the need to implement and control all terms (linear, quadratic, and $U$-term) of the Hamiltonian (\ref{LLHamiltonian}). 

The results can be extended to a launching scenario. To be specific, we shall consider 
the double well,  a paradigmatic quantum model that has been used, for example, 
to study and control some of the most fundamental quantum effects, like interference or tunneling. 
With the advent of ultracold-atom-based technology, it also finds applications in metrology, sensors,
and the implementation of basic operations for quantum information processing, 
like separation or recombination of ions \cite{Palmero2015}, as well as Fock state creation \cite{Martinez-Garaot2016}, 
and multiplexing/demultiplexing vibrational modes \cite{Martinez-Garaot2013,Martinez-Garaot2016tesis}. 
Here, we explore the possibility of using it for differential launching of vibrational  modes. 

\begin{figure}[t]
\begin{center}
\includegraphics[width=0.323 \linewidth]{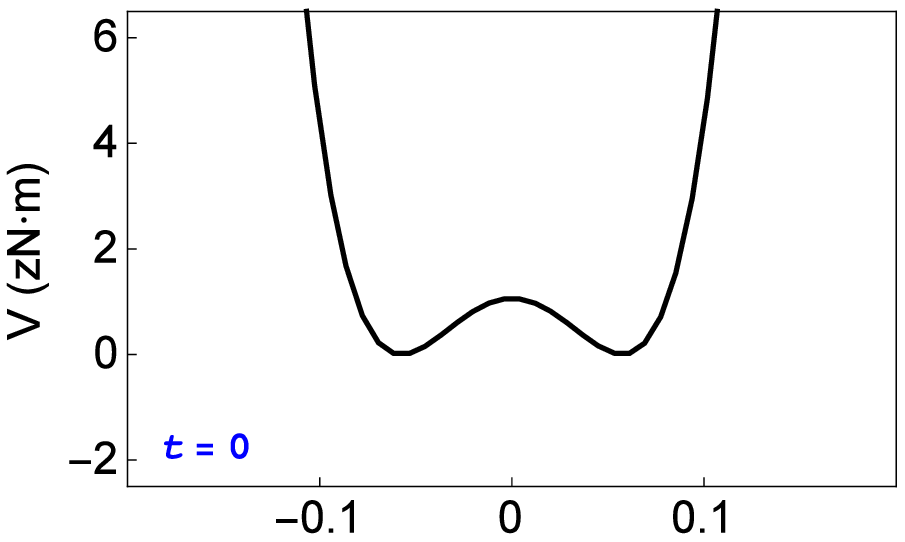}
\includegraphics[width=0.3 \linewidth]{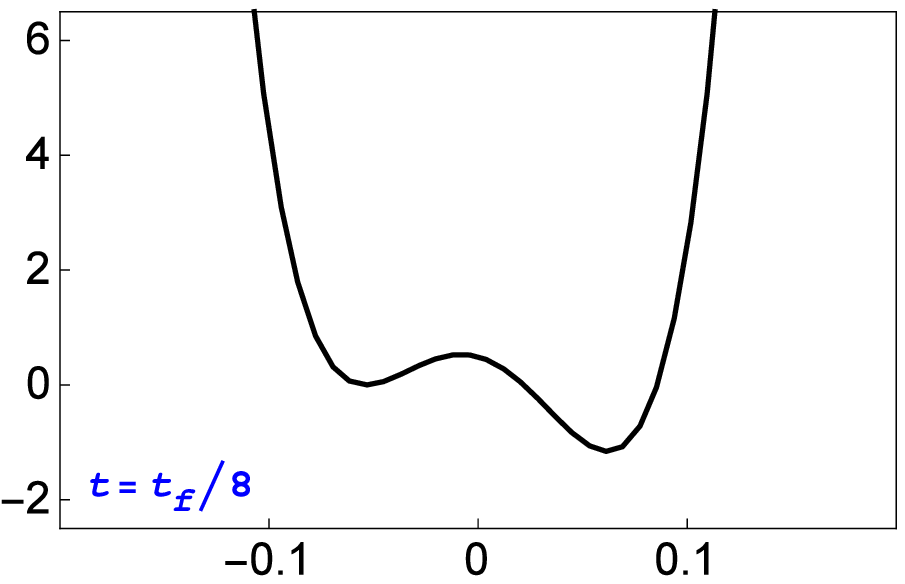}
\includegraphics[width=0.3 \linewidth]{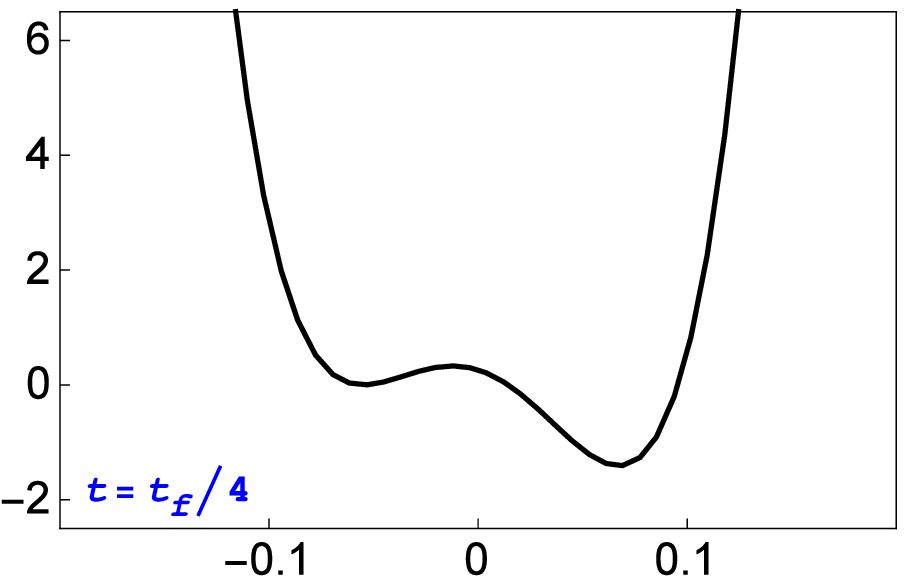}
\includegraphics[width=0.323 \linewidth]{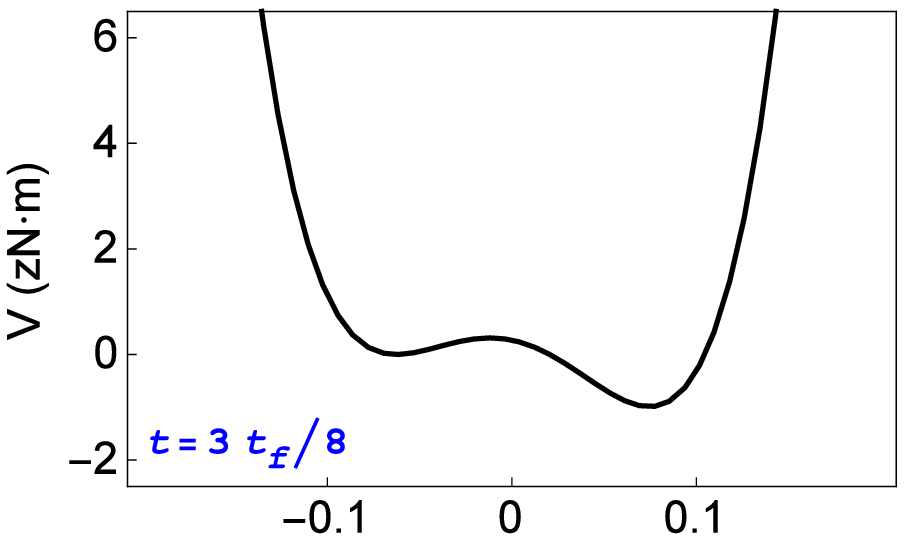}
\includegraphics[width=0.3 \linewidth]{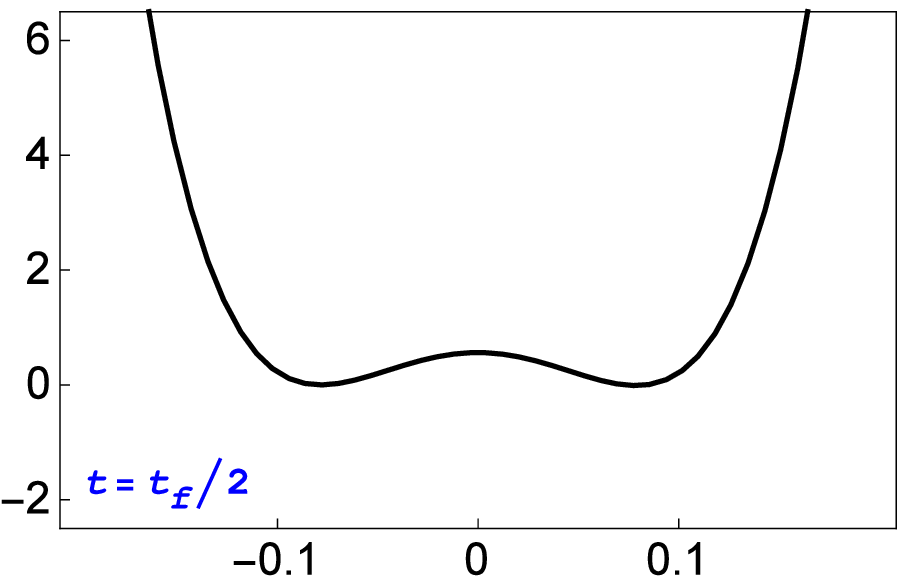}
\includegraphics[width=0.3 \linewidth]{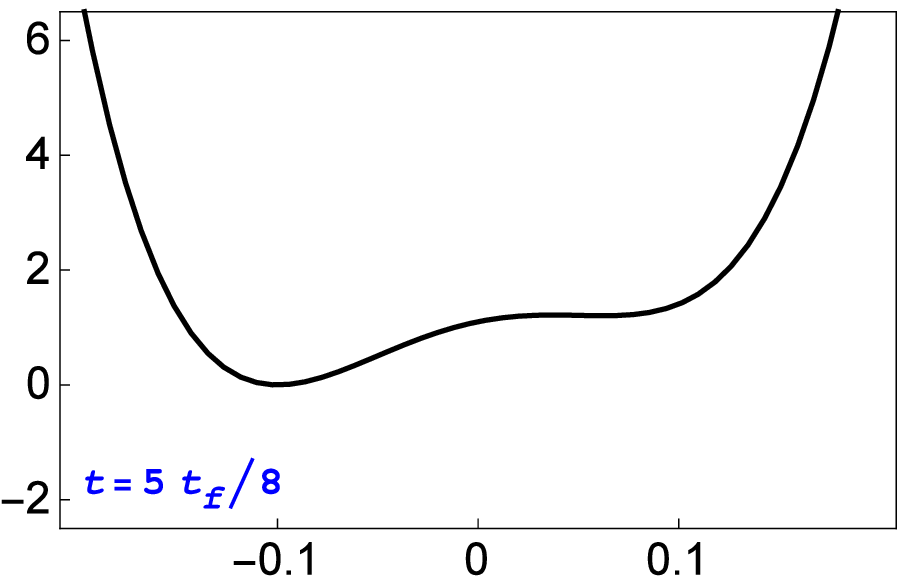}
\includegraphics[width=0.322 \linewidth]{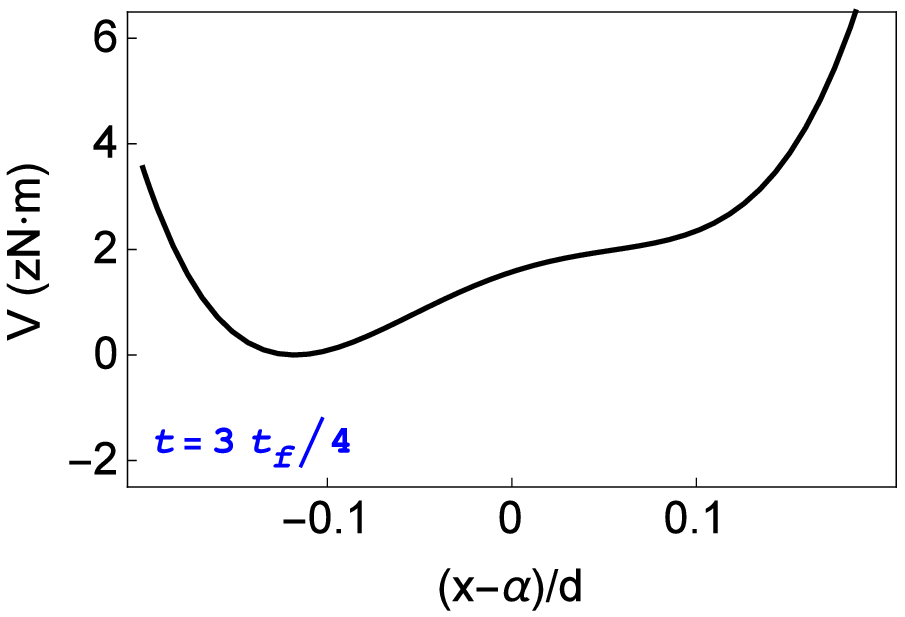}
\includegraphics[width=0.3 \linewidth]{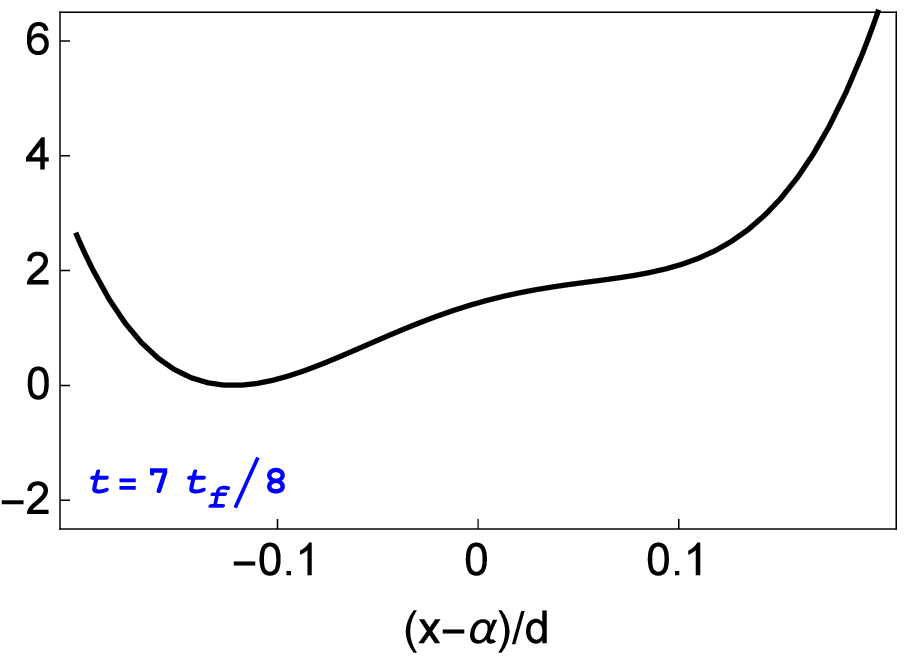}
\includegraphics[width=0.3 \linewidth]{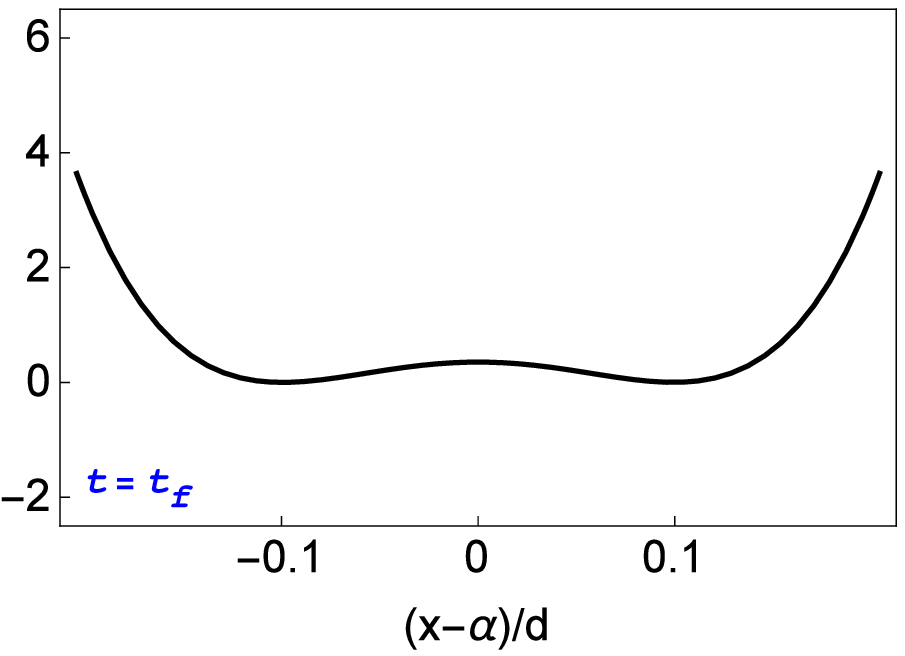}
\caption{\label{potential_snapshot}(Color online)
Time evolution of the shape of the launching double-well potential with velocities $v_f=10$ m/s and $\epsilon=2/s$. Each snapshot has been vertically displaced, without affecting the dynamics of the system, so that the minimum of the left well always lies at zero potential.
The parameters used are $\lambda=-8.7$ pN/m, $\beta=5.2$ mN/m$^3$, $\mu=86.4$ zN, $d=370$ $\mu$m, $\gamma=\sqrt{3}$ and $t_f=1$ $\mu$s. Even though not appreciated by the naked eye in the scale of the figures, the initial and final left wells are deeper than the right wells. }
\end{center}
\end{figure}

\begin{figure}[t]
\begin{center}
\includegraphics[width=0.5 \linewidth]{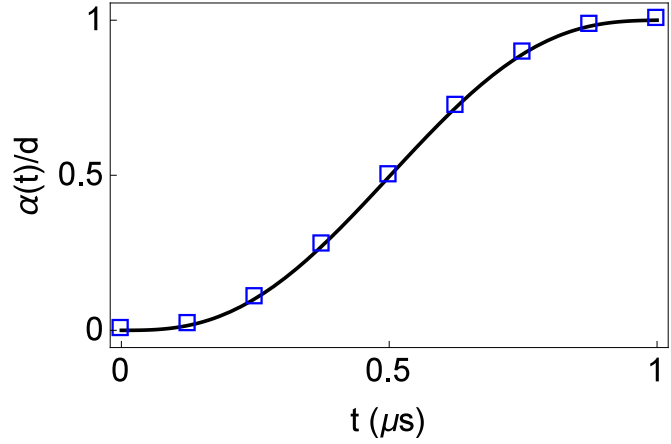}
\caption{\label{doublewell_trajectory}Scaled trajectory $\alpha$, Eq. (\ref{alphat}), of the center of the trap in a double well launching protocol with parameters $\lambda=-4.7$ pN/m, $\beta=5.2$ mN/m$^3$, $\mu=86.4$ zN, $d=370$ $\mu$m, $\gamma=\sqrt{3}$ and $t_f=1$ $\mu$s, and velocities
$v_f=10$ m/s and $\epsilon=2/s$. Blue rectangles mark the points of the trajectory in which a snapshot of the potential is depicted in Fig. \ref{potential_snapshot}.}
\end{center}
\end{figure}

We set $U$ (in $\sigma := \frac{x-\alpha}{\rho}$ space) as
\beq
U\left(\sigma \right)= \beta \sigma^4+\lambda \sigma^2+ \mu \sigma,
\label{dwell}
\eeq
where $\beta$, $\lambda$ and $\mu$ are constant parameters. 
$\beta$, is positive and $\lambda$ negative so that we have indeed  a double well. 
The linear term produces a  bias between the wells.
The condition \cite{Martinez-Garaot2015}
$
| \mu | \ll \frac{4\sqrt{2}}{3} \sqrt{-\frac{\lambda^3}{\beta}}
$
enables us to approximate $U(\sigma)$ as the sum of two harmonic potentials with minima at \cite{Martinez-Garaot2015}
\beq
\label{minimadoublewell}
\sigma_{\pm}(t) = \pm \frac 1 {\sqrt{2}}\sqrt{-\frac{\lambda}{\beta}}+\frac{\mu}{4\lambda}
\eeq
%
%Under this harmonic approximation, the effective potential at final time [see Eq. (\ref{potentialU})] is given by
%
%\beq
%V(x,t) \approx \frac{1}{2} m \Omega^2(x-x_+)^2+\frac{1}{2}m\Omega^2(x-x_-)^2,
%\eeq
%
in $\sigma$-space, and effective angular frequency  
%\cite{Martinez-Garaot2015}
%
\beq
\label{approxfreq}
\Omega  = 2 \sqrt{-\frac{\lambda}{m}}. 
\eeq
Limiting the linear  coefficient  as $|\mu|<\hbar(2\beta/m)^{1/2}$, 
the first excited and ground states lie in different wells \cite{Martinez-Garaot2013}.
We want to implement a protocol with a nonzero final expansion velocity, 
such that the effective launching velocities for  ground and first excited states are different
so that they separate further. 
We choose the boundary conditions for the auxiliary functions in Eqs. (\ref{bc1}) and (\ref{bc3}) 
and for the first derivatives   
\beqa
\label{bc2new}
\dot{\alpha}(0)&=&0,\; \dot{\alpha}(t_f)=v_0,
\nonumber\\
\dot{\rho}(0)&=&0,\; \dot{\rho}(t_f)=\epsilon. 
\eeqa
%
%, which will respectively be those in Eqs. (\ref{bc4}) and (\ref{bc6}).
Here the boundary conditions for the third derivatives [Eqs. (\ref{bc5}) and (\ref{bc7})] are not necessary.  
% because now $\alpha$ and $\rho$ 
%are laboratory parameters that control our resultant potential at boundary times. 
%That is, this potential will move and expand by the same parameters that move and expand the wave function, 
%so we do not need additional parameters to adjust its final velocity to be the same one of the packet.
With these conditions, using fifth-order polynomial ansatzes, the auxiliary functions are finally given by
\beqa
\alpha(t) &=& 2(5d-2t_f v_0)s^3+(-15d+7t_f v_0)s^4+3(2d-t_f v_0)s^5,
\label{alphat}
\\
\rho(t) &=& 1+ 2(-5+5\gamma-2t_f \epsilon)s^3+(15-15\gamma+7 t_f \epsilon)s^4
%\nonumber\\
+3(-2+2\gamma-t_f \epsilon)s^5.
\eeqa
These parameters directly give us the evolution of the potential term $\rho^{-2}U[(x-\alpha)/\rho]$. 
The auxiliary harmonic and linear terms in the total Hamiltonian (\ref{LLHamiltonian})  are found by  substituting $\alpha$ and $\rho$   
in Eqs. (\ref{auxiliaryharmonic}) and (\ref{auxiliarylinear}), respectively.
The resulting potential (the sum of the three potential terms in Eq. (\ref{LLHamiltonian})) is depicted in Fig. \ref{potential_snapshot} as a function of $(x-\alpha)/d$,
with $\alpha$ depicted 
in Fig. \ref{doublewell_trajectory}.

% % % % % % % % % % % % % % % % % % % % % % % % % % % % % % % % % % % % % % % % %
% % % % % % % % % % % % % % % % % % % % % % % % % % % % % % % % % % % % % % % %
% % % % % % % % % % % % % % % % % % % % % % % % % % % % % % % % % % % % % % % % % % %
% % % % % % % % % % % % % % % % % % % % % % % % % % % % % % % % % % % % % % % % % % %

For this evolution, we can calculate the average final velocity of the ground states in each well, and the final dispersion,
\beq
\langle v_\pm \rangle = v_0 + \epsilon \left(\frac{\mu}{4\lambda} \pm \frac{1}{\sqrt{2}}\sqrt{-\frac{\lambda}{\beta}} \right),\;\;\;\;\;\; 
\label{avv}
%\eeq
%
%and the final dispersion 
%
%\beq
\Delta v = \sqrt{\frac{\hbar}{4 \sqrt{-m\lambda}}\left( \epsilon^2-\frac{4\lambda}{m\gamma^2}\right)},
%\label{deltav}
\eeq
which is the same in both wells, as the effective frequency is also equal.  
Details of  these calculations are displayed in Methods.
Choosing the parameters so that  
$
\langle v_+ \rangle - \langle v_- \rangle > 2\Delta v,
$
guarantees that the wave packets of each well will never overlap.
\section*{Discussion}
In this paper, we have used the invariant-based inverse-engineering method to design shortcuts to adiabaticity for nonrigid driven transport and launching. 
Shortcuts for a harmonic trap are designed first, and then the analysis is extended to an arbitrary trapping potential. 
%For both cases, we studied not only the simple shuttling, but also the launching process.
Compared to rigid transport\cite{Torrontegui2011}, nonrigid transport 
requieres a more demanding manipulation, but it also provides a wider range of control opportunities, for example to 
achieve  narrow final velocity distributions in a launching process, suitable for accurate ion implantation or low-energy scattering experiments.  
A further example is the possibility to launch the ground states of each well in a double well with different velocities.
In a previous work \cite{Martinez-Garaot2013} processes to separate the ground and the first-excited states of a harmonic trap into different wells of 
a biased double well using STA were described.
The processes discussed here can be applied to different systems such as neutral atoms in optical traps, or 
classical mechanical oscillators, for which, mutatis mutandis, most of the results  apply.    
%We have showed here that it is
%
%
%
\section*{Methods}
\subsection*{Unitary diplacement and dilatation transformations}\label{unitransfor} 
First, we prove that given an arbitrary unitary transformation $U$, the transformed invariant 
$I'=UIU^\dagger$ is an invariant of  the effective  Hamiltonian $H'=UHU^\dagger + i \hbar \frac{\partial U}{\partial t} U^\dagger$.
Their commutator  is given by
\beq
[I',H'] = [UHU^\dagger + i \hbar \frac{\partial U}{\partial t} U^\dagger,UIU^\dagger] = U [H,I] U^\dagger + i \hbar \frac{\partial U}{\partial t} I U^\dagger -i \hbar UIU^\dagger \frac {\partial U}{\partial t} U^\dagger, 
\eeq
and the invariance condition [see Eq. (\ref{invequation})] for the transformed operators is satisfied, 
\beq
i \hbar\frac{\partial I'}{\partial t} - [I',H']= U\left( i \hbar \frac{\partial I}{\partial t} - [H,I] \right) U^\dagger =0.
\eeq
%

%To prove that the expanding transport mode in Eq. (\ref{eigenvecinv}) is an eigenstate of the Hamiltonian in Eq. (\ref{LLpotential}) in a comoving and coexpanding frame of reference, we follow procedure in [CITA PhysRevA.56.4300]. We go from the time dependent Hamiltonian which supports the invariant in Eq. (\ref{invariant}) to a static version, for which solutions are known in pronciple, by performing unitary transformations. Then, from these wave functions, we reach the solutions to the time dependent problem by undoing the previous transformations.
Now we introduce the specific unitary time-dependent operator $U=U_{d_2}\,U_{d_1}\,U_{p}\,U_{x}$.
%\beq
%\phi'_n(x)=U \, \phi_n(x,t),
%\eeq
%
%The transformation we apply is  the product of four unitary operators [CITA Lohe 2009]:
Operators $U_{d_1}$ and $U_{d_2}$ perform a time-dependent dilatation, and $U_{x}$ and $U_{p}$ a time-dependent translation in space and momentum, and are given by \cite{Lohe2009}
\beqa
U_{d_1}&=&e^{-\frac{i m \dot{\rho}}{2 \hbar \rho} x^2}; \hspace{1.5cm}
U_{d_2}=e^{\frac{i \ln \rho}{2 \hbar}(px+xp)}; \nonumber \\
U_{p}&=&e^{-\frac{i m \dot{\alpha}}\hbar x}; \hspace{1.7cm}
U_{x}=e^{\frac{i \alpha} \hbar x}.
\eeqa
%
%Under this transformation, the invariant in Eq. (\ref{invariant}) is mapped into
%
%\beq
%\label{transfeigenveceq}
%I'\,\psi'_n = \lambda_n \,\psi'_n,
%\eeq
%
%With $I'=UIU^\dagger$ and $\psi'_n=U\psi_n$. $I$ is given by Eq. (\ref{invariant}) and we assume that $\psi_n$ is ignored. Here, we fisrt solve Eq. (\ref{transfeigenveceq}) and then do the inverse transformation on the wave-funtion to find $\psi_n$ \cite{Pedrosa1997}. Opertion $I'=UIU^{\dagger}$ is detailed step by step:
%
In the comoving and coexpanding frame defined by this transformation, the new invariant
\beqa
\label{Iprime}
I' = U\, I \,U^\dagger = 
%\frac 1 {2m} \left[ \rho \left(p - m\dot{\alpha}\right) - m \dot{\rho} x\right]^2 + \frac12 m \omega_0\left(\frac x \rho \right)^2+U\left(\frac x \rho \right),\nonumber\\
%U_p \, U_x \, I \, U_x^\dagger \, U_p^\dagger = \frac 1{2m}\left [\rho p - m\dot{\rho}x\right]^2+\frac12 m \omega_0 \left(\frac x \rho \right)^2 + U \left( \frac x \rho \right),\nonumber\\
%U_{d1}\,U_p\,U_x\,I\,U_x^\dagger\,U_p^\dagger\,U_{d1}^\dagger= \frac1{2m}(\rho p)^2 + \frac12 m \omega_0 \left( \frac x \rho\right)^2 + U \left( \frac x \rho \right),\nonumber\\
U_{d_2}\,U_{d_1}\,U_p\,U_x\,I\,U_x^\dagger\,U_p^\dagger\,U_{d_1}^\dagger\,U_{d_2}^\dagger= \frac1{2m}p^2 + \frac12 m \omega_0 x^2 + U \left(x  \right),
\eeqa
becomes time independent \cite{Pedrosa1997}. Note that $I'$ has the same form of the Hamiltonian   in Eq. (\ref{auxschrodinger}) and therefore, the 
eigenstates of $I'$ are given by  $\phi_n(x)$. 
The inverse transformation  acting on $\phi_n$ provides the time dependent eigenvectors of $I(t)$ 
in Eq. (\ref{eigenvecinv}),
\beq
\psi_n(x,t)=U^\dagger \phi_n(x)=U_x^\dagger \, U_p^\dagger \, U_{d_1}^\dagger \, U_{d_2}^\dagger \, \phi'_n(x)
%\eeq
%
%Step by step:
%
%\beqa
%U_{d2}^\dagger\, \phi_n(x) = U_{d2}=e^{-\frac{i log \rho}{2 \hbar}(px+xp)} \phi_n(x) = \frac 1 {\sqrt{\rho}} \, \phi_n\left( \frac x \rho \right),\nonumber\\
%U_{d1}^\dagger \, U_{d2}^\dagger \, \phi_n(x)=e^{\frac{i m \dot{\rho}}{2 \hbar \rho} x^2} \frac 1 {\sqrt{\rho}} \, \phi_n\left( \frac x \rho \right),\nonumber\\
%U_p^\dagger\,U_{d1}^\dagger \, U_{d2}^\dagger \, \phi_n(x)=e^{\frac{im}\hbar\left( \frac{\dot{\rho}}{2\rho} x^2 \, + \, \dot{\alpha}x \right)}\frac 1 {\sqrt{\rho}} \, \phi_n\left( \frac x \rho \right),\nonumber\\
%U_x^\dagger \, U_p^\dagger\,U_{d1}^\dagger \, U_{d2}^\dagger \, \phi_n(x) =  e^{-\frac{i \alpha p}\hbar}e^{\frac{im}\hbar\left( \frac{\dot{\rho}}{2\rho} x^2 \, + \, \dot{\alpha}x \right)}\frac 1 {\sqrt{\rho}} \, \phi_n\left( \frac x \rho \right)\nonumber\\
%= e^{\frac{im}\hbar \left( \frac{\dot{\rho}x^2}{2\rho}+\dot{\alpha}x-\frac{\alpha \dot{\rho}x}\rho\right)}e^{-\frac{i \alpha p}\hbar}\frac 1 {\sqrt{\rho}} \, \phi_n\left( \frac x \rho \right)\nonumber\\
= e^{\frac{im}{\hbar} \left[ \dot{\rho}x^2/2\rho + (\dot{\alpha}\rho - \alpha \dot{\rho})x/\rho \right]} \frac{1}{\sqrt{\rho}} \phi_n \left( \frac{x-\alpha}{\rho} \right).
\eeq
%
%We obtain the same form as in Eq. (\ref{eigenvecinv}), thereby proving that, in a comoving and coexpanding reference frame, expanding transport modes in Eq. (\ref{eigenvecinv}) are eigenstates of the Hamiltonian in Eq. (\ref{LLHamiltonian}) at all times.
%
The Hamiltonian in the comoving and coexpanding frame is 
\beq
H' = UHU^\dagger + i \hbar \frac{\partial U}{\partial t} U^\dagger = \frac 1 {\rho^2} \left( \frac1{2m}p^2 + \frac12 m \omega_0 x^2 + U (x) \right) + \frac m 2 \left( \frac{\ddot{\rho}\alpha^2}{\rho} - \dot{\alpha} ^2 \right) - m \ddot{\alpha} \alpha,
\eeq
which, up to global terms that depend only on time, is proportional to the transformed invariant (\ref{Iprime}), so they commute at all times and thereby, share eigenstates at all times.

Note that the noninertial frame considered is comoving with $\alpha$, which is the center of the term $\rho^{-2}U[(x-\alpha)/\rho]$, 
but not necessarily the center of the harmonic potential $\frac{m}{2}\omega^2(x-x_0)^2$ in Eq. (\ref{Hamiltonian}) when $U=0$. 
However, the boundary conditions are set, see Eq. (\ref{bc5}),  so that indeed the frames moving with $\alpha$ and $x_0$ coincide
at boundary times $t_b=0, t_f$, as $\alpha(t_b)=x_0(t_b)$, and $\dot{\alpha}(t_b)=\dot{x}_0(t_b)$.  Similarly 
Eq. (\ref{bc7}) implies that the coexpanding frame depending on $\rho$ agrees with the one defined by the scaling factor
$\rho_{trap}=\sqrt{\omega_0/\omega_f}$ 
associated 
with the expansion of the trap, $\rho(t_b)=\rho_{trap}(t_b)$, and $\dot{\rho}(t_b)=\dot{\rho}_{trap}(t_b)$.      
\subsection*{Average velocity and dispersion in a double well}\label{AppendixC}
Here, we consider the  Hamiltonian in Eq. (\ref{LLHamiltonian}) with $U(\sigma)$ given a double well,   Eq. (\ref{dwell}), where ground and first-excited states lie in different wells and may be approximated 
by ground states of corresponding harmonic oscillators centered in $\sigma_\pm$ [see Eq. (\ref{minimadoublewell})], and effective angular frequency $\Omega$ [see Eq. (\ref{approxfreq})]. 
If the initial state is either the ground or first-excited state,  the dynamical state of the system is in either case
\beq
\label{statedoublewell}
\psi^\pm(x,t) = e^{\frac{im}{\hbar} \left[ \dot{\rho}x^2/2\rho + (\dot{\alpha}\rho - \alpha \dot{\rho})x/\rho \right]}\, \frac{1}{\rho^{1/2}}\, \phi_0^\pm,
%\left( \frac{x-\alpha}{\rho} - \sigma_\pm \right)%,
\eeq
where $\phi_0^\pm=\bigg(\frac{m\Omega}{\pi\hbar}\bigg)^{1/4}e^{-\frac{m\Omega}{2\hbar}\left(\frac{x-\alpha}{\rho} - \sigma_\pm \right)^2}H_0\Bigg[\sqrt{\frac{m\Omega}{\hbar}} \hspace{0.2cm}\left(\frac{x-\alpha}{\rho} - \sigma_\pm \right)\Bigg].$
%
%\beq
%\phi_0=\bigg(\frac{m\Omega}{\pi\hbar}\bigg)^{1/4}e^{-\frac{m\Omega}{2\hbar}\left(\frac{x-\alpha}{\rho} - \sigma_\pm \right)^2}H_0\Bigg[\sqrt{\frac{m\Omega}{\hbar}} \hspace{0.2cm}\left(\frac{x-\alpha}{\rho} - \sigma_\pm \right)\Bigg].
%\eeq
%

%%The average velocity and the dispersion are given by
%
%%\beqa
%%\langle v_\pm \rangle = - \frac{i\hbar}{m} \int (\psi^\pm)^\ast \partial_x \psi^\pm \,dx,\;\;\; 
%\\
%%\Delta v_\pm= \sqrt{\langle v^2_\pm \rangle - \langle v_\pm \rangle ^2},
%%\eeqa
%
%%where $\langle v^2_\pm \rangle = -\frac{\hbar^2}{m^2} \int (\psi^\pm)^\ast \partial_x^2\psi^\pm \, dx$.
%\beq
%\langle v^2 \rangle = -\frac{\hbar^2}{m^2} \int \psi^\ast \partial_x^2\psi \, dx.
%\eeq
%

Using  standard properties of Hermite polynomials the average of the velocity and its square are found to be 
\beqa
\langle v_\pm \rangle &=& - \frac{i\hbar}{m} \int (\psi^\pm)^\ast \partial_x \psi^\pm \,dx = \dot{\alpha} + \dot{\rho}\,\sigma_\pm,
\label{avvap}
\\
%\eeq
%
%Using again the properties of Hermite polynomials, 
%We only collect terms proportional to $\psi_n$, the only ones contributing to the integral,
%
%\beqa
%\partial_x^2\psi_n
%&=&\left(\frac{im\dot{\alpha}}{\hbar}+\frac{im\dot{\rho}}{\hbar}\sigma_\pm \right)^2\psi_n+\left(-\frac{m^2\dot{\rho}^2}{4\hbar^2}\frac{2n\hbar}{m\Omega}-\frac{1}{4\rho^2}\frac{2nm\Omega}{\hbar}\right)\psi_n\nonumber\\
%&+&\left(-\frac{m^2\dot{\rho}^2}{4\hbar^2}\frac{2(n+1)\hbar}{m\Omega}-\frac{1}{4\rho^2}\frac{2(n+1)m\Omega}{\hbar}\right)\psi_n\nonumber\\
%= \left[ -\frac{m^2}{\hbar^2}(\dot{\alpha}+\dot{\rho}\sigma_\pm)^2-\frac{m\dot{\rho}^2}{\hbar\Omega}\left(n+\frac{1}{2}\right)-\frac{m\Omega}{\rho^2\hbar}\left(n+\frac{1}{2}\right)\right]\psi_n+...
%\eeqa
%
%Doing the integral, rearranging terms, and setting $n=0$, 
%\beq
%\langle v^2 \rangle = -\frac{\hbar^2}{m^2} \int \psi^\ast \partial_x^2\psi \, dx=(\dot{\alpha}+\dot{\rho}x_\pm)^2+\frac{\hbar(2n+1)}{2m\Omega}\left(\dot{\rho}^2+\frac{\Omega^2}{\rho^2}\right).
\langle v^2_\pm \rangle &=& -\frac{\hbar^2}{m^2} \int (\psi^\pm)^\ast \partial_x^2\psi^\pm \, dx=(\dot{\alpha}+\dot{\rho}\sigma_\pm)^2+\frac{\hbar}{2m\Omega}\left(\dot{\rho}^2+\frac{\Omega^2}{\rho^2}\right).
\eeqa
Finally, the dispersion, common to both wells,  is given by
\beq
\Delta v= \Delta v_\pm=\sqrt{\langle v_\pm^2 \rangle - \langle v_\pm \rangle ^2}=\sqrt{\frac{\hbar}{2m\Omega}\left(\dot{\rho}^2+\frac{\Omega^2}{\rho^2}\right)}.
\label{deltavap}
\eeq
Equation (\ref{avv})  follows by substituting in Eqs. (\ref{avvap}) and (\ref{deltavap}) 
the expressions for $\sigma_\pm$ and $\Omega$, Eqs. (\ref{minimadoublewell}) and (\ref{approxfreq}), 
and the final values of the auxilary functions and their derivatives in Eqs. (\ref{bc1}) and (\ref{bc2new}).   
\bibliography{Bibliography}

\section*{Acknowledgements}
We thank G. C. Hegerfeldt for discussions. 
This work was partially supported by the Basque Government (Grant IT986-16), 
and Grant FIS2015-67161-P (MINECO/FEDER,UE).
%, and the program UFI 11/55 of UPV/EHU. 
M.P. and S.M.-G. acknowledge fellowships by UPV/EHU.
\section*{Author contributions statement}
%
%Must include all authors, identified by initials, for example:
All authors conceived the work, discussed the results, and reviewed the manuscript,  A.T. conducted the calculations. 
\section*{Additional information}
%
%To include, in this order: \textbf{Accession codes} (where applicable); 
%\textbf{Competing financial interests}: 
The authors declare no competing financial interests.  
%(mandatory statement). 

%The corresponding author is responsible for submitting a \href{http://www.nature.com/srep/policies/index.html#competing}{competing financial interests statement} on behalf of all authors of the paper. This statement must be included in the submitted article file.

%\begin{figure}[ht]
%\centering
%\includegraphics[width=\linewidth]{stream}
%\caption{Legend (350 words max). Example legend text.}
%\label{fig:stream}
%\end{figure}

%\begin{table}[ht]
%\centering
%\begin{tabular}{|l|l|l|}
%\hline
%Condition & n & p \\
%\hline
%A & 5 & 0.1 \\
%\hline
%B & 10 & 0.01 \\
%\hline
%\end{tabular}
%\caption{\label{tab:example}Legend (350 words max). Example legend text.}
%\end{table}
%
%Figures and tables can be referenced in LaTeX using the ref command, e.g. Figure \ref{fig:stream} and Table \ref{tab:example}.

\end{document}